\documentclass[superscriptaddress,secnumarabic,
amssymb,amsmath,nobibnotes,aps,prd,showkeys,showpacs,nofootinbib]{revtex4}%
\usepackage{graphicx}
\usepackage{epsf}
\usepackage{bm}
\usepackage{amsmath}
\usepackage{amsfonts}
\usepackage{amssymb}
\usepackage{epstopdf}
\usepackage{natbib}
\usepackage{color}
\usepackage{float}
\providecommand{\U}[1]{\protect\rule{.1in}{.1in}}

\newcommand{\be}{\begin{equation}}
\newcommand{\ee}{\end{equation}}

\newcommand{\mincir}{\raise
-3.truept\hbox{\rlap{\hbox{$\sim$}}\raise4.truept\hbox{$<$}\ }}
\newcommand{\magcir}{\raise
-3.truept\hbox{\rlap{\hbox{$\sim$}}\raise4.truept\hbox{$>$}\ }}

\newtheorem{remark}{Remark}[section]

\begin{document}


\title{Gravitational particle production of superheavy massive particles in Quintessential Inflation II: $\alpha$-attractors}

\author{Llibert Arest\'e Sal\'o}
\email{l.arestesalo@qmul.ac.uk} 
\affiliation{School of Mathematical Sciences, Queen Mary University of London, Mile End Road, London, E1 4NS, United Kingdom}

\author{Jaume  de Haro}
\email{jaime.haro@upc.edu}
\affiliation{Departament de Matem\`atiques, Universitat Polit\`ecnica de Catalunya, Diagonal 647, 08028 Barcelona, Spain}



\begin{abstract}
We compute   the gravitational production of conformally coupled superheavy  particles during the  phase transition from the end of inflation to the beginning of  kination for  $\alpha$-attractors potentials in the context of Quintessential Inflation ($\alpha$-QI), showing that the
 maximum value of the reheating temperature, independently of the value of the parameter $\alpha$, is near $10^9$ GeV. This result, which contradicts  the usual belief that the reheating via  the production of superheavy massive particles leads  to an inefficient reheating temperature, is due to the fact that in our numerical calculations we take into account the contribution of the large wavelength modes to the reheating temperature, which never happens in analytical calculations where only  ultraviolet modes are considered.
\end{abstract}

\vspace{0.5cm}

 \par
\maketitle
\section{Introduction}

The so-called {\it gravitational particle production}
  \cite{Parker,gmm,ford,Zeldovich} of superheavy particles conformally coupled with gravity, which was applied to standard inflation (potentials with a deep well) in \cite{kolb, kolb1,Birrell1, ema}, is one of the mechanisms used to reheat the universe in  scenarios containing a period of inflation. This also happens in Quintessential Inflation (QI) to match the inflationary period with the usual hot Big Bang universe
\cite{guth}.
  However, the gravitational reheating in QI is normally applied to very light fields 
   \cite{Spokoiny, pv, A} and only in few papers, which deal with toy non-smooth   models as the Peebles-Vilenkin one, it is applied to massive particles
   \cite{H, ha, hap18, J,hashiba, Hashiba}. On the other hand, regarding smooth QI potentials, particle creation 
    has to be analytically studied using the complex WKB approximation \cite{hashiba1}, whose
   effective application is limited to the creation of particles by parabolic potentials \cite{kofman}.

\

In the present  work we continue with our study of particle production by smooth potentials started in \cite{ah}, now focusing on a smooth exponential  potential coming from $\alpha$-attractors in Quintessential Inflation 
($\alpha$-QI) \cite{ vardayan, K, benisty3}. 
Since the $\alpha$-attractors come from supergravity theories   containing particles with only gravitational interactions, the late-time decay of these relics  may jeopardize the success of the standard BBN \cite{lindley}. To solve this problem one has to consider 
 sufficiently low reheating temperature (of the order of $10^9$ GeV or less) \cite{eln}. On the contrary,  a lower bound for the reheating temperature comes from the fact that the
radiation-dominated  era occurs before the Big Bang Nucleosynthesis (BBN) epoch, which takes place in the  $1$ MeV  regime \cite{gkr}.

\

Coming back to the gravitational production of superheavy particles,  
we will  use the well-known Hamiltonian diagonalization method (see \cite{gmmbook} for a review), based on the computation of  the time dependent 
$\beta$-Bogoliubov coefficient  which encodes the polarization effects 
and also the real particles created during the phase transition. Fortunately,  
these polarization effects disappear when the universe evolves adiabatically, which happens soon after the beginning of kination, allowing its numerical calculation. Thus, in order to   calculate the energy density of the produced particles, one can safely use the 
square modulus of the $\beta$-Bogoliubov coefficient after the beginning of kination, whose numerical value is, for the relevant modes that contribute to the reheating, of the order of $10^{-9}-10^{-10}$ depending on the superheavy masses which in our simulations are of the order of $10^{15}-10^{17}$ GeV.


\



Finally, once one has  the value of the $\beta$-Bogoliubov coefficients,  one can calculate the value of the energy density of the superheavy particles, which must decay into lighter ones before or after the end of the kination phase to form a relativistic plasma. In the former case the reheating temperature is greater than $10^6$ GeV and in the second one its maximum value is around $10^9$ GeV, which shows that the gravitational production of superheavy particles is a very efficient mechanism to reheat our universe. 

\

Throughout  the manuscript we  use natural units, i.e., 
 $\hbar=c=k_B=1$ and the reduced Planck's mass is denoted by $M_{pl}\equiv \frac{1}{\sqrt{8\pi G}}\cong 2.44\times 10^{18}$ GeV.

\section{Particle creation of superheavy  particles conformally coupled to gravity}
\label{sec-particle-creation}

We consider a superheavy quantum field $\chi$ conformally coupled with gravity. In order that the polarization effects due to this quantum field do not affect the dynamics of the scalar field responsible for Quintessential Inflation, the mass of the quantum field, namely $m_{\chi}$,  must be greater than the Hubble rate, in fact the condition  $H\ll m_{\chi}$ has to be satisfied (see \cite{Felder} to take a look at the problems associated to quantum  light  fields). Therefore, since the most accepted idea is that inflation starts at GUT scales, where the temperature is around $10^{16}$ GeV, then using the Stefan-Boltzmann law $\rho=\frac{\pi^2}{30}g_* T^4$, where the effective number of degrees of freedom for the Standard model is $106.75$, for a flat FLRW spacetime at the beginning of inflation one has $H\cong 2\times 10^{13}$ GeV. This is the reason why we will chose $m_{\chi}\sim 10^{15}$ GeV or greater.

\

To calculate the energy density of the produced particles,  we will use the well-known {\it diagonalization method}, based on the Bogoliubov coefficients, which in the conformally coupled case must satisfy the first order system of differential equations (see \cite{gmmbook} for a detailed discussion)
\begin{eqnarray}\label{Bogoliubovequation}
\left\{ \begin{array}{ccc}
\alpha_k'(\tau) &=& \frac{\omega_k'(\tau)}{2\omega_k(\tau)}e^{2i\int^{\tau} \omega_k(\bar\tau)d\bar\tau}\beta_k(\tau)\\
\beta_k'(\tau) &=& \frac{\omega_k'(\tau)}{2\omega_k(\tau)}e^{-2i\int^{\tau}\omega_k(\bar\tau)d\bar\tau}\alpha_k(\tau),\end{array}\right.
\end{eqnarray}
where the time dependent frequency is denoted by  $\omega_k(\tau)=\sqrt{k^2+m_{\chi}^2a^2(\tau)}$ and  $\tau$ is the conformal time.

\

Finally, in terms of the $\beta$-Bogoliubov coefficient, the vacuum energy density of the  $\chi$-field is given by \cite{ah}
\begin{eqnarray}\label{vacuum-energy1}
\rho_{\chi}(\tau)= \frac{1}{2\pi^2a^4(\tau)}\int_0^{\infty} k^2\omega_k(\tau)|\beta_k(\tau)|^2 dk.
\end{eqnarray}
\

\subsection{ Gravitational particle creation by $\alpha$-attractors in the context of Quintessential Inflation}

\begin{figure}[ht]
\centering
\includegraphics[width=0.4\textwidth]{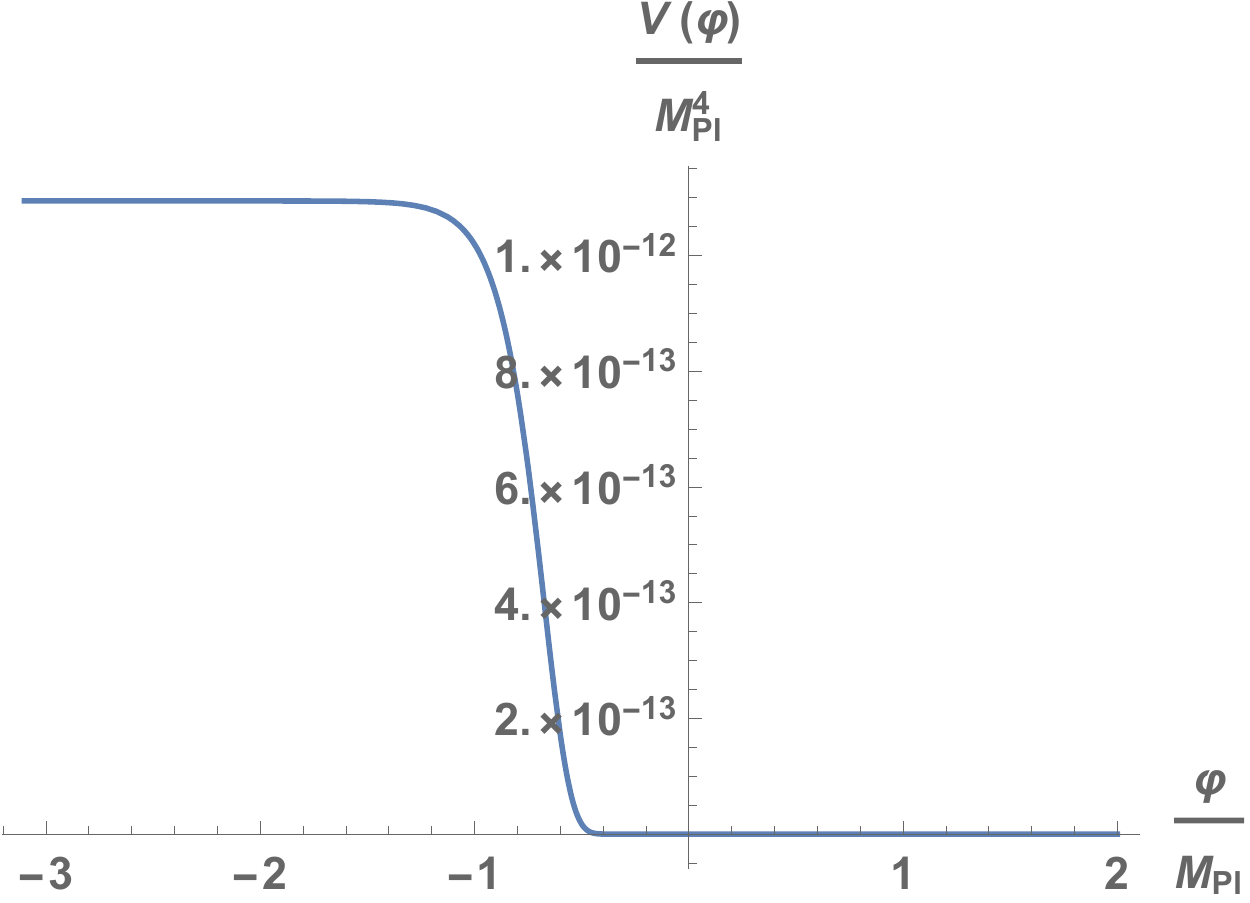}
\caption{{Plot of the Exponential $\alpha$-attractor potential, for $\alpha\sim 10^{-2}$, $n\sim 10^2$ and $\lambda\sim 10^{-66}$.}}
\label{fig:attr}
\end{figure}

In the present work, the potential that we will consider is an 
exponential $\alpha$-attractor in Quintessential Inflation, plotted in Figure \ref{fig:attr} and given by 
\begin{eqnarray}\label{alpha}
V(\varphi)=\lambda M_{pl}^4e^{-n\tanh\left(\frac{\varphi}{\sqrt{6\alpha}M_{pl}} \right)},
\end{eqnarray}
where $\lambda$, $\alpha$ and $n$ are dimensionless parameters, whose relation is in order to match with the current observation data the following one (see  \cite{benisty3} for details):
\begin{eqnarray}\label{parameters}
\frac{\lambda}{\alpha}e^{n}\sim 10^{-10} \qquad \mbox{and}\qquad \lambda e^{-n}\sim 10^{-120}.
\end{eqnarray}



First of all, in order to calculate the energy density of the produced particles one has to integrate numerically the conservation equation for the inflaton field, namely
\begin{eqnarray}\label{conservation}
\ddot{\varphi}+3H\dot{\varphi}+V_{\varphi}=0,
\end{eqnarray}
where $H=\frac{1}{\sqrt{3}M_{pl}}\sqrt{\frac{\dot{\varphi}^2}{2}+V(\varphi)  }$, with initial conditions (the value of the scalar field and its first derivative) during inflation. Since the slow-roll regime is an attractor, one only has to take initial conditions in the basin of attraction of the slow-roll solution, for example,
$\varphi=\varphi_*$ and $\dot{\varphi}=-\frac{V_{\varphi}(\varphi_*)}{3H_*}$, where the ``star" denotes that the quantities are evaluated
at the horizon crossing.

\

Once one has obtained the evolution of the scalar field and in particular the evolution of the Hubble rate, 
one can compute the evolution of the scalar factor, whose value at the horizon crossing we have chosen to be equal to $1$.


\

\begin{figure}[ht]
\includegraphics[width=0.48\textwidth]{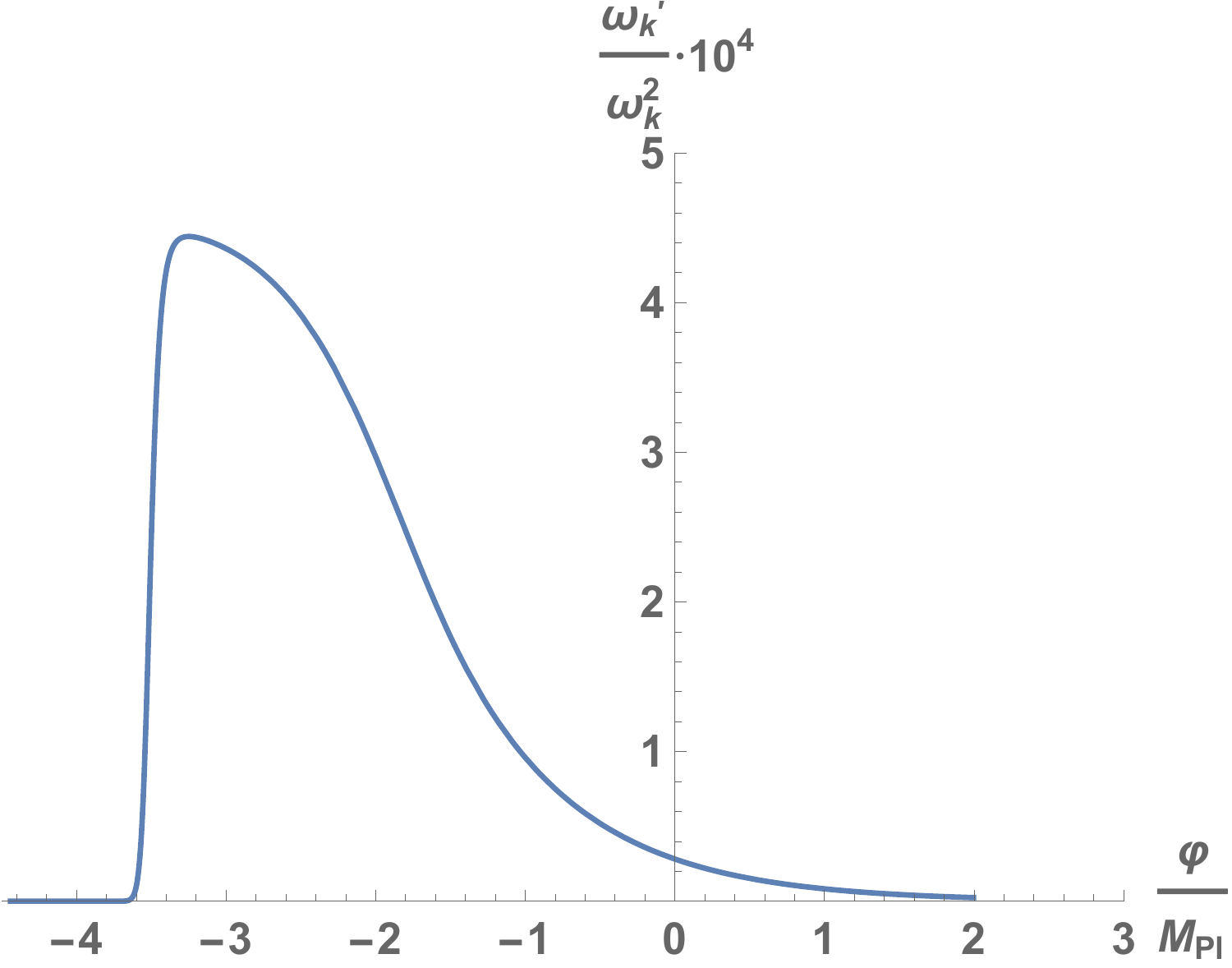}
\includegraphics[width=0.5\textwidth]{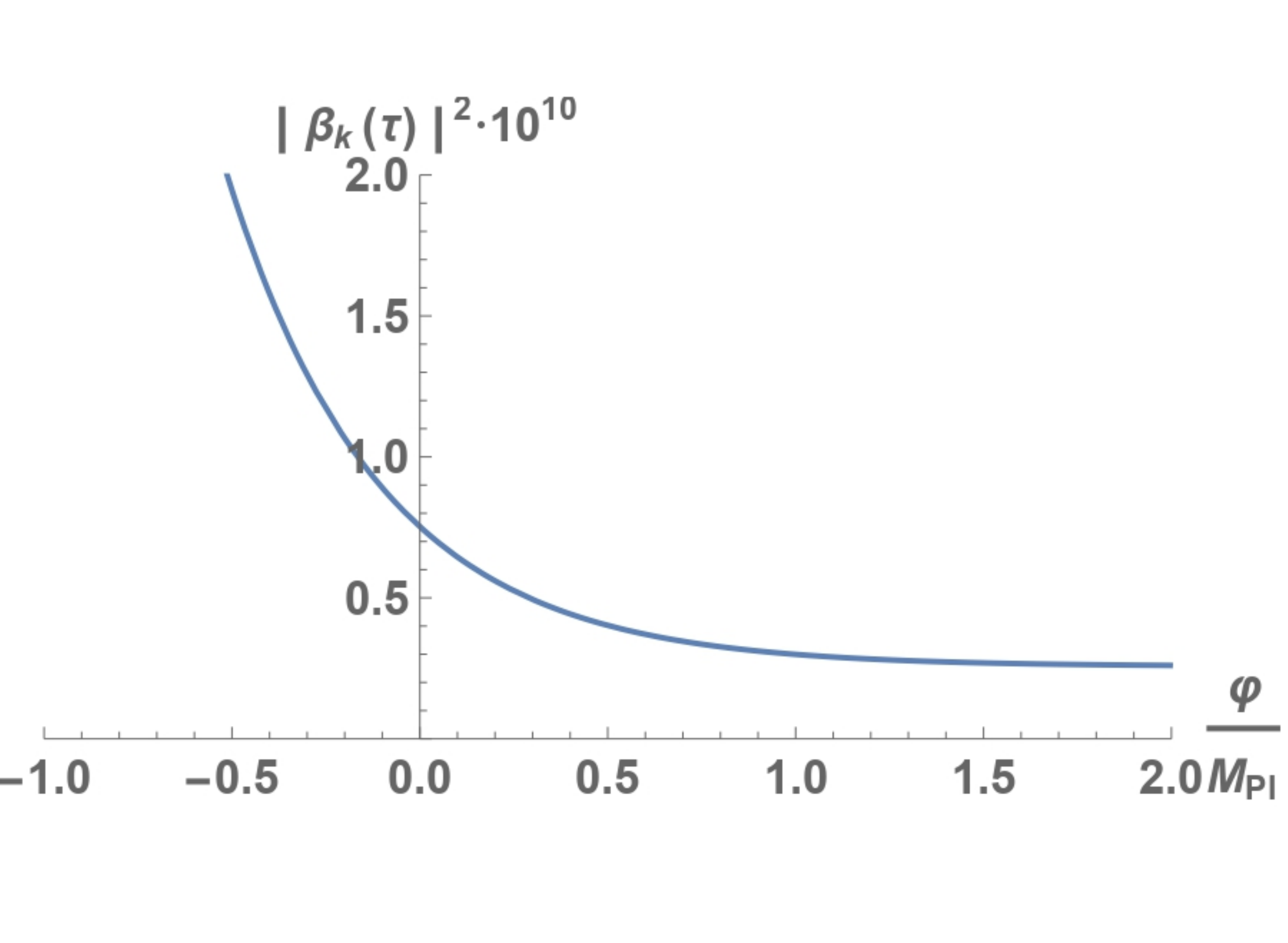}
\caption{Adiabatic evolution (left) and evolution of the $\beta$-Bogoliubov coefficient (right) for a heavy field with mass $m_{\chi}\sim 10^{16}$ GeV, when $\alpha\sim 10^{-1}$. Here we have used the value $k=a_{kin}H_{kin}$,  which is in the range 
${k}\lesssim a_{kin}m_{\chi}$, where we have observed that particles are produced.}
\label{fig:bogo}
\end{figure}

From the evolution of the scale factor, 
 we can see on the left-hand side of Figure \ref{fig:bogo} that a spike appears in the plot of the quantity $\omega_k'/\omega_k^2$ during the phase transition from the end of inflation to the beginning of kination, that is, at that moment when the adiabatic evolution is broken and particles are gravitationally produced (see the right-hand side of Figure \ref{fig:bogo}).


\

Then,  we have numerically solved equation (\ref{Bogoliubovequation}), 
with initial conditions  $\alpha_k(\tau_*)=1$ and $\beta_k(\tau_*)=0$ at the horizon crossing (there were neither particles nor polarization effects at that moment because during the slow-roll regime the derivatives of the Hubble rate are negligible compared with the powers of $H$, i.e.,  the system is in the adiabatic regime).

\

\begin{figure}[ht]
\includegraphics[width=0.6\textwidth]{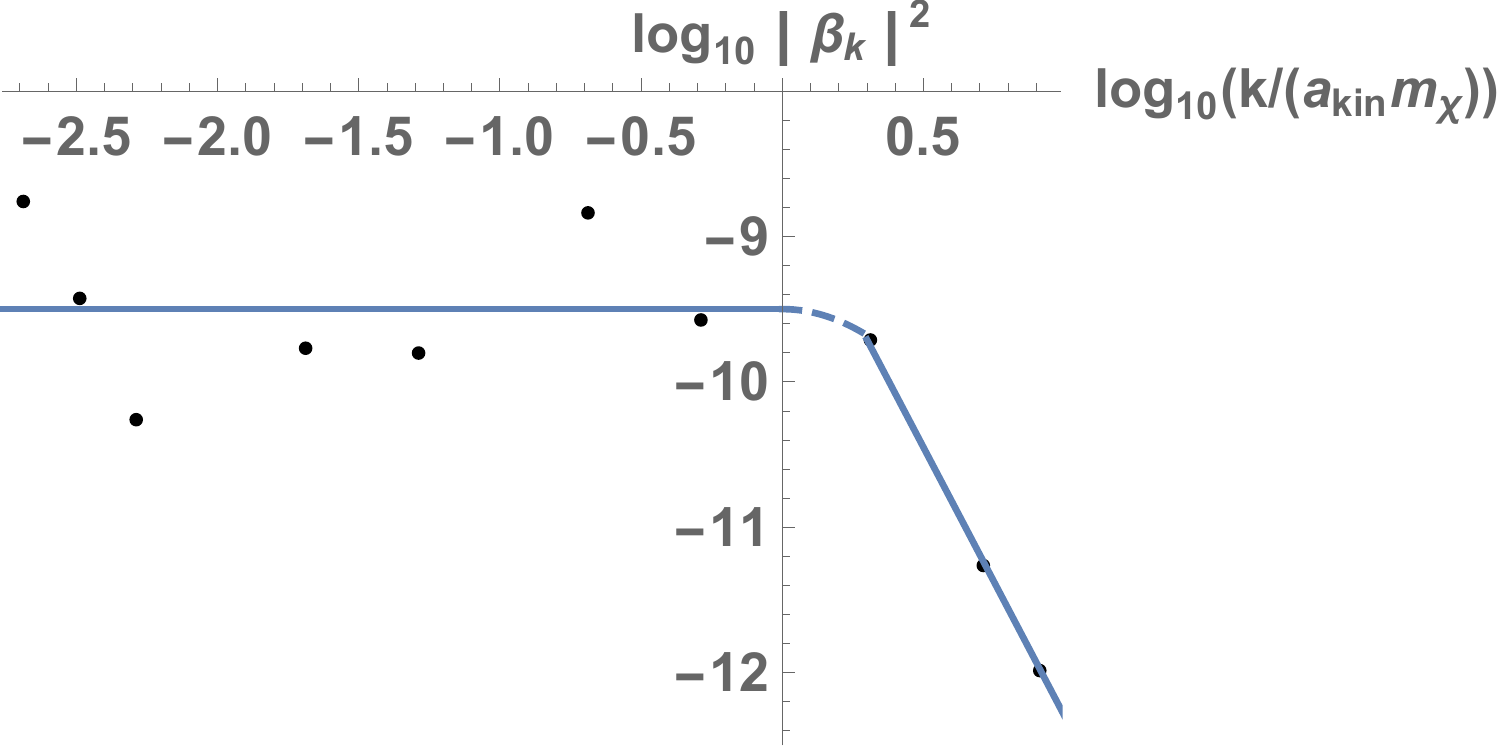}
\caption{Plot of the logarithm of $|\beta_k|^2$,  as a function of $k$, for a heavy field with mass $m_{\chi}\sim 10^{15}$ GeV and $\alpha=0.1$.  }
\label{fig:betaksm15a01}
\end{figure}

\begin{figure}[ht]
\centering
\includegraphics[width=0.52\textwidth]{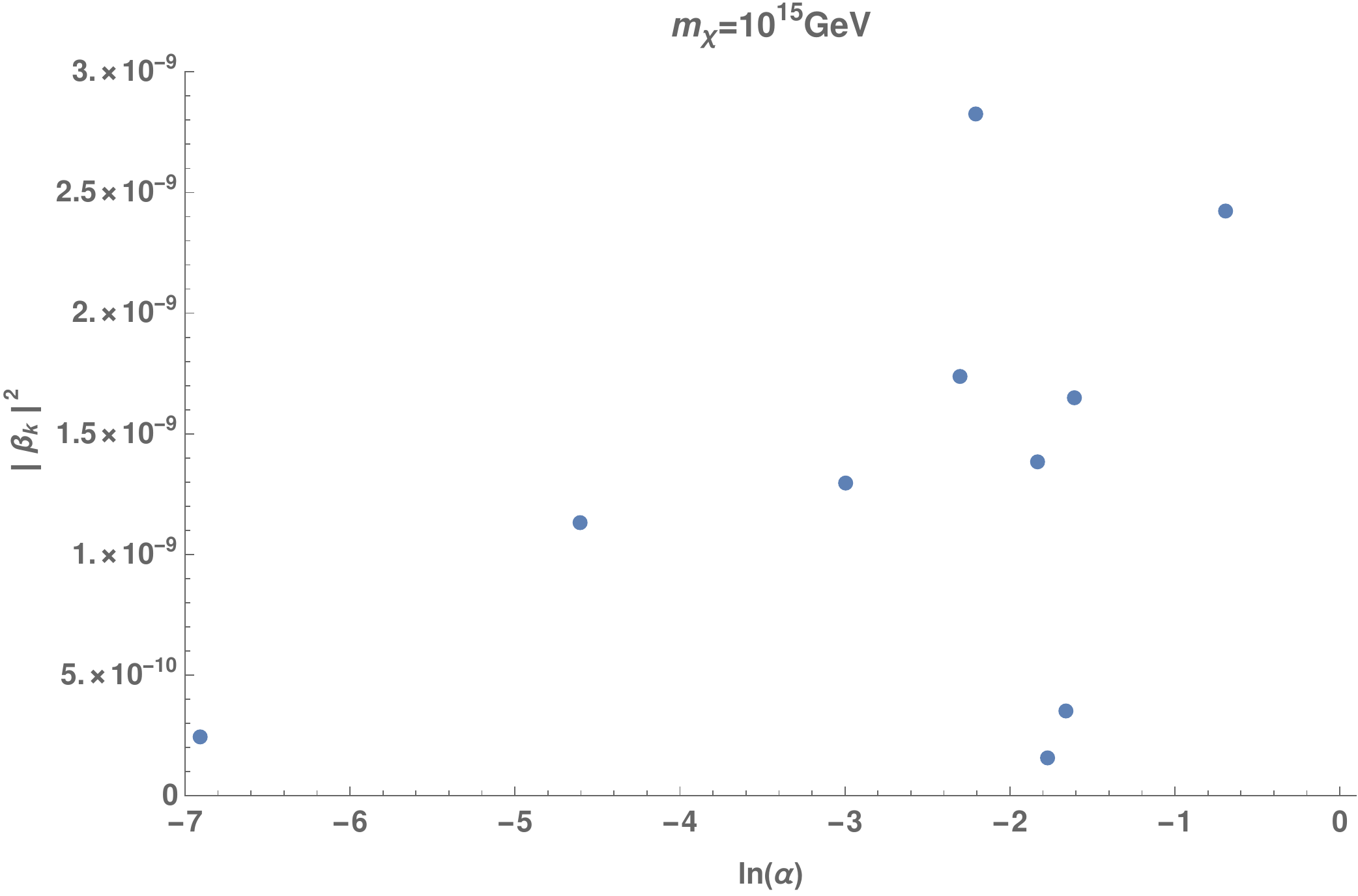}
\includegraphics[width=0.495\textwidth]{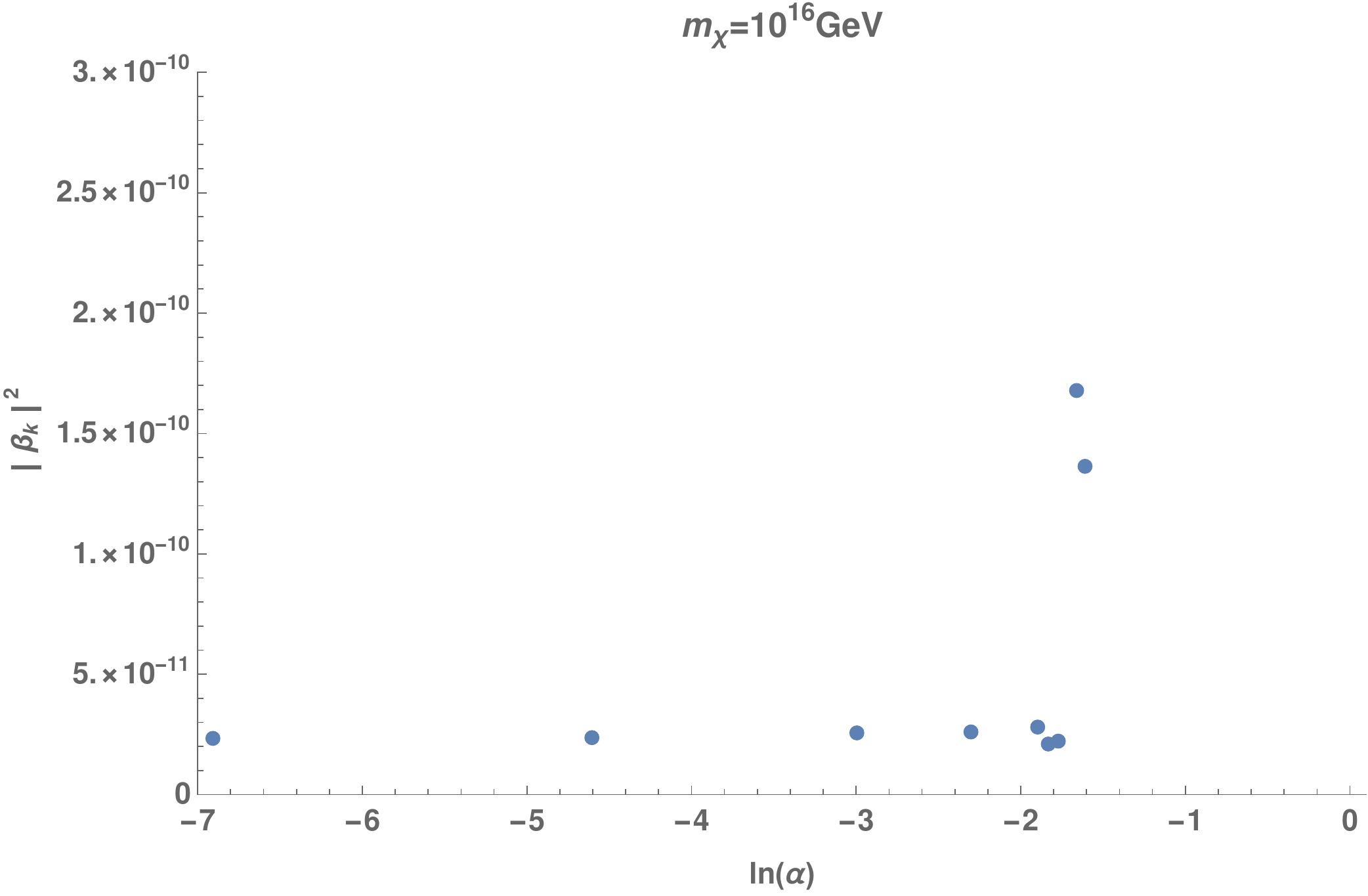}
\includegraphics[width=0.495\textwidth]{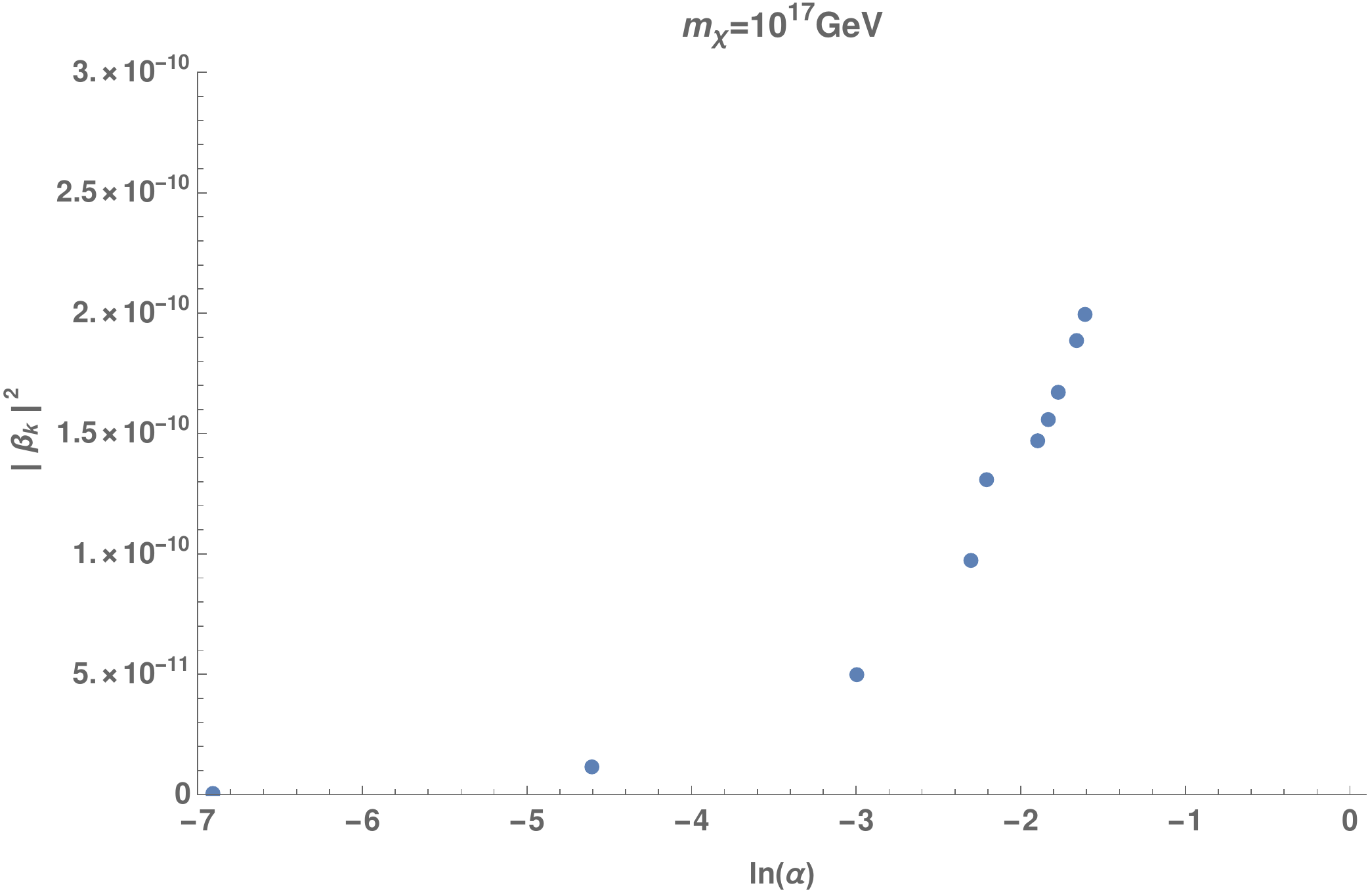}
\caption{Plot of the square modulus of the  $\beta$-Bogoliubov coefficient,  as a function of $\alpha$, for a heavy field with masses $m_{\chi}\sim 10^{15}$, $10^{16}$ and $10^{17}$ GeV.}
\label{fig:betams}
\end{figure}

For the value $k=a_{kin}H_{kin}$, we obtain in Figure \ref{fig:bogo} that $|\beta_k(\tau)|^2$ stabilizes soon to a non-zero value after the beginning of kination, containing only particle production effects. We have numerically done the calculations for masses  $m_{\chi}\cong 10^{15}-10^{17}$ GeV and for a large range of values of $\alpha$. We have obtained that the relevant modes that contribute significantly  to the particle production are in the range 
$0\lesssim k\lesssim a_{kin}m_{\chi}$ (see Figure \ref{fig:betaksm15a01}), leading to
 values of $|\beta_k|^2$ of order $10^{-9}$ for $m_{\chi}\sim 10^{15}$ GeV and values of $|\beta_k|^2$ of order $10^{-10}$ for $m_{\chi}\sim 10^{16}-10^{17}$ GeV, as one can see in Figure \ref{fig:betams}.


\



\



\


\






Next, introducing these values of the $\beta$-Bogoliubov coefficient in the energy density 
(\ref{vacuum-energy1}) and taking into account that the modes that contribute to the energy density satisfy 
$0\lesssim k\lesssim a_{kin}m_{\chi}$ and lead practically to the same value of the $\beta$-Bogoliubov coefficient, one can safely do the approximation that after the beginning of kination $\omega_k(\tau)\cong m_{\chi}a(\tau)$, obtaining
\begin{eqnarray}
\rho_{\chi}(\tau)\cong \frac{m_{\chi}^4}{6\pi^2}
|\beta_k|^2
\left(\frac{a_{kin}}{a(\tau)}\right)^3\sim
\left\{\begin{array}{ccc}
   10^{-11}m_{\chi}^4 \left(\frac{a_{kin}}{a(\tau)}\right)^3 & \mbox{when} & m_{\chi}\sim 10^{15} \mbox{ GeV} \\
   & &\\
10^{-12}m_{\chi}^4 \left(\frac{a_{kin}}{a(\tau)}\right)^3     & \mbox{when} & m_{\chi}\sim 10^{16}-10^{17} \mbox{ GeV},
\end{array}\right.
\end{eqnarray}
for $\tau>\tau_{kin}$.

\



A final remark is in order: In \cite{hashiba,ema}, the authors infer through numerical calculations using toy models that, for the relevant modes, the square of the $\beta$-Bogoliubov coefficient must decay as $e^{-\kappa m_{\chi}/H_{inf}}$, where $H_{inf}$ is the scale of inflation and $\kappa$ a dimensionless factor of order 1. However, throughout all the calculation the mass of the superheavy field is of the same order or less than the scale of inflation. So, in practice, for the relevant models the authors obtain in all numerical simulations
$|\beta_k|^2\geq 10^{-7}$.  

\

In our case we have chosen masses which are many orders greater than the scale of inflation.
Effectively, 
since the power spectrum of scalar perturbations $P_{\zeta}=\frac{H_*^2}{8\pi^2\epsilon_*M_{pl}^2}\sim 2\times 10^{-9}$ (here, once again, the star denotes that the quantities are evaluated at the horizon crossing)
leads to  $H_*\sim 4\times 10^{-4} \sqrt{\epsilon_*}M_{pl}$ -with $\epsilon_*=\frac{3\alpha}{16}(1-n_s)^2$ for the case of  $\alpha$-attractors, choosing for the spectral index  its central value $n_s=0.9649$ (see \cite{planck})-, one gets the following  scale of inflation,
\begin{eqnarray}\label{Hubble}
H_*\sim 1.5\sqrt{\alpha}\times 10^{13} \mbox{ GeV},
\end{eqnarray}
which is minimum two orders less than $10^{15}$ GeV. 

\

In summary, there is gravitational production of superheavy particles, with the square of the $\beta$-Bogoliubov, for the relevant modes, of the order of $10^{-9}-10^{-10}$, but for superheavy masses the production is not suppressed by an abnormally small factor as
$e^{-\kappa m_{\chi}/H_{inf}}$.

\section{The reheating process}
\label{sec-reheating}

After the production of the superheavy particles,  they have to decay into lighter ones which after the thermalization process form a relativistic plasma that depicts our hot universe. Then, this decay could occur before the end of kination (when the energy density of the inflaton becomes of the same order than the one of the $\chi$-field), or after its end.



\subsection{Decay before the end of kination}
\label{sebsec1-reheating}

 In this case,
 the energy density of the background, i.e. the one of the inflaton field,  and the one of the relativistic plasma, when the decay is finished, 
that is
 when ${\Gamma}\sim H_{dec}=H_{kin}\left(\frac{{a}_{kin}}{a_{dec}} \right)^3$,
   will be
\begin{eqnarray}\label{LQIrho}
\rho_{\varphi, dec}=3{\Gamma}^2M_{pl}^2\qquad 
 \mbox{and}  \qquad
 \rho_{\chi,dec}=
\rho_{\chi,kin}\left(\frac{{a}_{kin}}{a_{dec}} \right)^3\cong \frac{m_{\chi}^4}{6\pi^2}
|\beta_k|^2
\frac{\Gamma}{H_{kin}}.\end{eqnarray}

Imposing that the end of the decay precedes the end of kination, which means $ \rho_{\chi,dec}\leq \rho_{\varphi, dec}$, 
and taking into account that it is after the beginning of the kination,  i.e.,  $\Gamma\leq H_{kin}\cong 6\times 10^{-7} M_{pl}$,
one gets
\begin{eqnarray}\label{bound}
\frac{1}{18\pi^2}
|\beta_k|^2
\frac{m_{\chi}^4}{H_{kin}M_{pl}^2}\leq \Gamma\leq H_{kin}.
\end{eqnarray}

\

Finally,  the reheating temperature, i.e., the temperature of the universe when the relativistic plasma in thermal equilibrium starts to dominate, 
which happens when $\rho_{\varphi, reh}\sim \rho_{\chi,reh}$, can be calculated as follows:
Since after the decay the evolution of the respective energy densities are given by
\begin{eqnarray}
\rho_{\varphi, reh}=\rho_{\varphi, dec}\left(\frac{a_{dec}}{a_{reh}}\right)^6 \qquad \mbox{and} \qquad
 \rho_{\chi,reh}= \rho_{\chi,dec} \left(\frac{a_{dec}}{a_{reh}}\right)^4, \end{eqnarray}
we will have
$\frac{ \rho_{\chi,dec}}{\rho_{\varphi,dec}}=\left(\frac{a_{dec}}{a_{reh}}\right)^2,$
and thus,
from the Stefan-Boltzmann law $\rho_{reh}=\frac{\pi^2 }{30}g_{reh}T_{reh}^4$, where $g_{reh}=106.75$ is the effective number of degrees of freedom for the Standard Model, the reheating temperature will be
\begin{eqnarray}\label{reheating1}
 T_{reh}=  \left(\frac{30}{\pi^2g_{reh}} \right)^{1/4}
 \rho_{\chi,reh}^{\frac{1}{4}} = 
 \left(\frac{30}{\pi^2g_{reh}} \right)^{1/4}
 \rho_{\chi,dec}^{\frac{1}{4}}
 \sqrt{\frac{\rho_{\chi,dec}}{\rho_{\varphi,dec}}} \nonumber\\
\cong \frac{1}{3\sqrt{2}\pi^{3/2}}\left(\frac{30}{\pi^2g_{reh}} \right)^{1/4} 
|\beta_k|^{3/2}
\left( \frac{H_{kin}}{6\Gamma}\right)^{1/4}
\frac{m_{\chi}^3}{M_{pl}^2H_{kin}}
M_{pl}.
\end{eqnarray}

 So, taking into account the bound (\ref{bound}),
the reheating temperature ranges between  
 \begin{eqnarray}
 \frac{1}{3\sqrt{2}\pi^2}\left(\frac{5}{g_{reh}}\right)^{1/4}
 |\beta_k|^{3/2}
\frac{m_{\chi}^3}{M_{pl}^2H_{kin}}
M_{pl} \leq T_{reh}\leq \nonumber\\  \frac{1}{3\sqrt{2}\pi^{3/2}}\left(\frac{90}{g_{reh}}\right)^{1/4}
|\beta_k|
\left(\frac{m_{\chi}}{M_{pl}}\right)^2\sqrt{\frac{M_{pl}}{H_{kin}}}M_{pl},
 \end{eqnarray}
which shows that $6\times 10^5$ GeV $\lesssim T_{reh}\lesssim 6\times 10^8$ GeV for $m_{\chi}\sim 10^{15}$ GeV and $T_{reh}\gtrsim 10^8$ GeV for $m_{\chi}\sim 10^{16}$ GeV. However, for masses of order $10^{17}$ GeV, the reheating temperature is some orders greater than $10^9$ GeV. This fact, as we have already explained in the Introduction, supposes a great problem in order to ensure the BBN success.
\

Therefore, to have a viable reheating temperature, we conclude that only superheavy particles with masses of the order $10^{15}-10^{16}$ GeV could decay before the end of kination. 

\



\subsection{Decay after the end of kination}

In the case that the decay of the $\chi$-field is after the end of kination (recall that kination ends when $\rho_{\varphi}\sim \rho_{\chi}$),
one  has to impose ${\Gamma}\leq H(\tau_{end})\equiv H_{end}$, where we have denoted by $\tau_{end}$ the time at which kination ends. Taking this into account, one has 
\begin{eqnarray}\label{31}
H^2_{end}=\frac{2\rho_{\varphi, end}}{3M_{pl}^2}
\end{eqnarray}
and \begin{eqnarray}  \rho_{\varphi, end}={\rho}_{\varphi,kin}\left( \frac{{a}_{kin}}{a_{end}} \right)^6=
\frac{ {\rho}_{\chi,kin}^2}{{\rho}_{\varphi,kin}},
\end{eqnarray}
where we have used that the kination ends when ${ {\rho}_{\chi,end}}={{\rho}_{\varphi, end}}$, meaning 
$\left({a}_{kin}/a_{end} \right)^3=
\frac{{\rho}_{\chi,kin}}{{\rho}_{\varphi,kin}}$. So, the condition ${\Gamma}\leq H_{end}$ leads to the bound 
\begin{eqnarray}\label{bound1}
\Gamma\leq \sqrt{\frac{2}{3}}\frac{ \rho_{\chi,kin}}{M_{pl}\sqrt{\rho_{\varphi,kin}}}\cong
\frac{\sqrt{2}}{18\pi^2}
|\beta_k|^2
\frac{m_{\chi}^4}{H_{kin}M_{pl}^2} .
\end{eqnarray}

\

 On the other hand, assuming once again instantaneous thermalization, the reheating temperature (i.e., the temperature of the universe when the thermalized plasma starts to dominate) will be obtained when all the superheavy particles decay, i.e. when $H\sim \Gamma$, obtaining
\begin{eqnarray}
T_{reh}=\left( \frac{30}{\pi^2 g_{reh}} \right)^{1/4}\rho_{\chi,dec}^{1/4}= \left( \frac{90}{\pi^2 g_{reh}} \right)^{1/4}\sqrt{{\Gamma}M_{pl}},
\end{eqnarray}
where we have used that, after  the end of the kination regime, the energy density of the produced particles dominates the one  of the inflaton field. 

\

Consequently,  since the BBN epoch occurs at the $1$ MeV regime, one can find that, in that case,  the reheating temperature is bounded by
\begin{eqnarray}
1 \mbox{ MeV}\leq T_{reh} \leq \frac{1}{3\pi} \left( \frac{45}{\pi^2 g_{reh}} \right)^{1/4}
|\beta_k|
\left(\frac{m_{\chi}}{M_{pl}}\right)^2\sqrt{\frac{M_{pl}}{H_{kin}}}M_{pl},
\end{eqnarray}
which for masses $m_{\chi}\sim 10^{15}-10^{17}$ GeV leads to  a reheating temperature in all the range of viable values $1$ MeV $\leq T_{reh}\leq 10^9$ GeV.

\



\section{ A simple model containing the Cosmological Constant}

A very  simple model comes from a lineal potential  with the Cosmological Constant denoted by $\Lambda$ \cite{vardayan},  $V(\phi)=\lambda(\phi+\sqrt{6\alpha})+\Lambda M_{pl}^2$, which in terms of the canonically unitary field $\varphi$, defined as  $\phi=\sqrt{6\alpha}\tanh \left(\frac{\varphi}{\sqrt{6\alpha}M_{pl}}\right)$, becomes
\begin{eqnarray}\label{cc}
V(\varphi)=\lambda \sqrt{6\alpha}\left( \tanh \left(\frac{\varphi}{\sqrt{6\alpha}M_{pl}}\right) +1 \right)+\Lambda M_{pl}^2.
\end{eqnarray}

To obtain the values of the parameters, first of all we calculate the main slow roll parameter
\begin{eqnarray}
\epsilon\cong \frac{1}{12\alpha \cosh^4\left(\frac{\varphi}{\sqrt{6\alpha}M_{pl}}\right)\left( \tanh \left(\frac{\varphi}{\sqrt{6\alpha}M_{pl}}\right) +1 \right)^2},
\end{eqnarray}
thus, in order that inflation ends, which happens when $\epsilon=1$, one needs that $\frac{1}{12\alpha}>1$. So, we will take  for example $\alpha\sim 10^{-2}$. 

\

To get the value of $\lambda$ we use the formula (\ref{Hubble}) and the fact that at the horizon crossing 
\begin{eqnarray}
H_*^2\cong \frac{V(\varphi_*)}{3M_{pl}^2}\cong \frac{2\lambda \sqrt{6\alpha}}{3M^2_{pl}}.
\end{eqnarray}
Combining both results one gets $\frac{\lambda}{\sqrt{\alpha}}\sim 2\times 10^{-11} M_{pl}^4$.

\

Finally, it is well-known that, to match with the current observational data, the Cosmological Constant $\Lambda$ must be of the order 
$10^{-120} M_{pl}^2$ because at the present time the current observational  data state that  $\Omega_{\Lambda}=\frac{\Lambda M_{pl}^2}{3H_0^2 M_{pl}^2}\cong 0.7$.

\

\begin{remark}
The potential (\ref{cc}) does not belong to the class of Quintessential Inflation potentials because to deal with the current cosmic acceleration it needs a Cosmological Constant. The difference with the potential (\ref{alpha}) is that this one contains two parameters, namely $\lambda$ and $n$, needed to unify the early and late-time acceleration. However, the potential $\lambda \sqrt{6\alpha}\left( \tanh \left(\frac{\varphi}{\sqrt{6\alpha}M_{pl}}\right) +1 \right)$ only contains the parameter $\lambda$ which is determined by the power spectrum of scalar perturbation. Then, to correctly depict the late-time acceleration one needs another parameter, in this case the Cosmological Constant.

\end{remark}

\begin{remark}
Note also that for the potential  (\ref{alpha}) the value of $\alpha$ is not restricted to be small as for the potential  (\ref{cc}), because for that potential the main slow roll parameter satisfies
$\epsilon=\frac{n^2}{12\alpha \cosh^4\left(\frac{\varphi}{\sqrt{6\alpha}M_{pl}}\right)}$, so in order that inflation ends one needs $\frac{n^2}{12\alpha}>1$, and since $n$ is of the order $10^2$ (see the equation (\ref{parameters}) and also \cite{benisty3} for a detailed discussion) this allows a large range of values of $\alpha$.
\end{remark}

\

The important point  is that  this model leads to the same results for particle production as the exponential model studied previously in detail, and thus, it leads to the same bounds for the reheating temperature.


\section{Conclusions}

In the present work we have numerically studied the gravitational particle production of superheavy particles conformally coupled to gravity in $\alpha$-Quintessential Inflation. To calculate  the energy density of the produced particles we have used the well-known diagonalisation method, where the key point is the calculation   of the time-dependent 
$\beta$-Bogoliubov coefficient. 

\

This coefficient encodes all the polarization effects  and the produced superheavy particles  during the phase transition from the end of inflation to the beginning of kination.
 Fortunately,  the polarization effects disappear soon after the  beginning of kination, which enables us to extract from it only the part which has to do with particle production. In fact, from the
 relevant modes that contribute to the particle production the square modulus of the $\beta$-Bogoliubov coefficient is of the order
 $10^{-9}-10^{-10}$, depending on the mass of superheavy particles.
 
 



\

Once these superheavy particles have been created, they must decay into lighter ones to form a relativistic plasma which eventually becomes dominant and matches with the hot Big Bang universe. Then, two different situations arise, namely when the decay occurs  before the end of the kination regime and when the decay occurs after the end of the kination regime. We have shown that
for both situations the maximum value of the reheating temperature is quite big, more or less around $10^9$ GeV, which demystifies the belief  that heavy masses suppress the particle production, thus leading to  an abnormally  low reheating temperature. What really happens is that  the main contribution of particle production is in  a long wavelength regime, which 
without a numerical calculation
is impossible to quantify.  This is the reason why   in many papers the production of superheavy particles is simply disregarded  
 because  analytically only ultraviolet effects can be calculated, but, as it is well-known, the ultraviolet modes do not contribute significantly  to the particle production.

\section*{Acknowledgments}
JdH is supported by grant MTM2017-84214-C2-1-P funded by MCIN/AEI/10.13039/501100011033 and by ``ERDF A way of making Europe'',
and  also in part by the Catalan Government 2017-SGR-247. L.A.S thanks the School of Mathematical Sciences (Queen Mary University of London) for the support provided.

\end{document}